\documentstyle{article}
\begin{document}
\title{Phase Space Distributions from Three-Port Couplers}
\author{Matteo G. A. Paris, Alexei V. Chizhov, and Ole Steuernagel\\
Arbeitsgruppe ``Nichtklassiche Strahlung"
der Max-Planck-Gesellschaft \\ Rudower Chaussee 5, 
12489 Berlin, Germany}
\date{}
\maketitle
\begin{abstract}
A wide class of phase space distributions of a single
mode radiation field is shown to be directly accessible
to measurement by linear symmetric three-port optical
couplers.
\end{abstract}
In a classical description of the harmonic oscillator the measurement
of some variable can be performed, at least in principle, with
arbitrary precision. In particular, a pair of conjugated variables
can jointly be measured, thus providing a unique correspondence
between states of the oscillator and points in the phase space of
complex amplitudes. This is no longer true in quantum
mechanics, to which classical phase space description cannot be
transferred \cite{wig}. Two main features have to be
considered: i) the uncertainty principle which prevents precise
identification of points in phase space and ii) the commutation
relations which lead to ordering rules for representing
physical variables in terms of operators~\cite{gla}.
Nevertheless, a phase space description of the harmonic oscillator
has been very fruitful in quantum optics and different
distribution functions have been introduced to describe various
dynamical processes. Some examples are laser theory, the study of the phase
properties of light~\cite{rip} and the quantum state
measurement problem~\cite{vog,par}.
\par
An important class of quantum phase space distributions are
the so-called generalized Wigner functions~\cite{gla}
\begin{equation}
W_s (\alpha,\bar\alpha ) = \int \frac{d^2\lambda}{\pi}
\; \chi_s (\lambda,\bar{\lambda})
\; e^{\bar\lambda \alpha -\lambda \bar\alpha}
\label{Wdf}\;,
\end{equation}
where 
\begin{equation}
 \chi_s (\lambda,\bar{\lambda}) =
\hbox{Tr}\left\{\hat\rho\;e^{\lambda a^{\dag} -\bar\lambda a+\frac{1}{2}s
|\lambda |^2}\right\}
\label{Xis}\;
\end{equation}
is the $s$-ordered characteristic function of the
field $\hat\rho$ and is related to different ordering
of the boson operators $a$, $a^{\dag}$ since the statistical average
$\langle \bar \alpha^k \alpha^l \rangle_s$ --~employing the quasi-probability
distribution $W_s (\alpha,\bar\alpha )$~--
provides the quantum expectation value of the $s$-ordered
operator product $\{ a^{\dag k} a^l \}_s$.
\\
For $s=1$ we get Glauber's $\cal P$-function~\cite{sud}, for $s=0$,
the distribution originally proposed by Wigner~\cite{wig} and
for $s=-1$ the Husimi $Q$-function~\cite{hus} which are
related to normal, symmetric and antinormal ordering of the boson
operators respectively. 
Such generalized Wigner functions can be sampled experimentally only
for parameter $s\leq -1$, since for greater values of $s$
negative or singular probabilities occur for some quantum states.
\par
A more general class of phase distributions has been
introduced in quantum optics~\cite{wod}. These are defined as
a convolution of
the Wigner function of the signal $W_{0|S}$ with that of the
probe $W_{0|P}$ and serve as a description of the signal in terms of
the probe:
\begin{equation}
K_{SP}(\alpha ,\bar{\alpha}) = \int_{\bf C} \frac{d^2 \beta}{\pi^2} \;
W_{0|S}(\alpha+\beta,\bar\alpha +\bar\beta ) \;
W_{0|P} (\beta,\bar\beta ) 
\label{Con}\; .
\end{equation}
After some calculation this can be shown to be equivalent to 
\begin{equation}
K_{SP}(\alpha ,\bar{\alpha}) = \frac{1}{\pi}\hbox{Tr}\left\{
\hat\rho_S\hat D(\alpha )\hat\rho_P\hat D^{\dag}(\alpha )
\right\}
\label{Tra}\; ,
\end{equation}
which represents $K_{SP}$ in terms of the overlap between the 
signal field $\hat\rho_S $ and the $\alpha$-displaced probe field $\hat\rho_P$; 
$\hat D(\alpha )=\exp\{\alpha a^{\dag}-
\bar{\alpha }a\}$ is the displacement operator. 
Phase space density as in Eq. (\ref{Con}) have been introduced
to account for the effect of the measuring apparatus
in a joint measurement of conjugated variables~\cite{wod}. 
More recently, they also have been used in entropic descriptions
of quantum mechanical states~\cite{buz}.
\par
In this paper we consider {\em triple coupler homodyning} of a single mode
radiation  field, that is, the use of a linear three-port coupler to couple 
the signal
beam with a strong local oscillator and also with a second probe field.
Upon using vacuum as the probe mode input this
leads to the measurement of Husimi's $Q$-function, a result which is a 
by-product of the present approach and is somewhat 
interesting since the measurement of the $Q$-function is commonly
associated with an $4 \times 4$ - port device~\cite{rip,leo}
(eight-port homodyne \cite{lai}) although we see here a $3 \times 3$ - port
is enough. For a general probe field we will find that triple coupler homodyning
allows access to all possible phase space density $K_{SP}$, 
provided the corresponding field $\hat\rho_P$ can be generated.
The use of a lower number of modes results in an easier 
and less noisy experimental 
implementation.
\par
A linear, symmetric three-port optical coupler  is a
generalization of the customary lossless symmetric beam splitter.
The three input modes $a_i$, $i=1,2,3$ are combined to form 3 output
modes $b_j$, $j=1,2,3$.
In analogy to lossless beam splitters which are described by unitary 
2$\times$2 matrices~\cite{cam},
any lossless symmetric triple coupler is characterized by a unitary
$3\times 3$ matrix~\cite{zap,zin} of the form
\begin{eqnarray}
{\bf T} = \frac{1}{\sqrt{3}}
\left(
\begin{array}{ccc}
1 & 1 & 1 \\
1 & \exp\{i\frac{2\pi}{3}\}  &  \exp\{-i\frac{2\pi}{3}\}  \\
1 & \exp\{-i\frac{2\pi}{3}\}  &  \exp\{i\frac{2\pi}{3}\}
\end{array}
\right)
\label{T3M}\;,
\end{eqnarray}
where each matrix element $T_{ij}$ represents the transmission amplitude
from the $i$-th input port to the $j$-th output port, that is 
$b_j = \sum_{k=1}^3 T_{jk} a_k$.
\par
Such devices have already been implemented in single-mode optical
fiber technology and commercial triple coupler have been available for
some time~\cite{she}. It is known~\cite{zpl} that any unitary
$M$-dimensional matrix can be factorized into a sequence of 2-dimensional
transformation. Thus any triple coupler can be implemented by discrete
optical components using 50:50 beam splitters and phase shifters only~\cite{zap}, 
which is schematically displayed in Fig.~\ref{f:TTT} together with
our notation for general linear three-port couplers.
Experimental realizations of triple couplers  has been reported for both cases, 
the passive elements case and the optical fiber one \cite{zap,zin}.
\par
Let us now consider the measurement scheme of Fig.~\ref{f:THD}.
The three input modes are mixed by a triple coupler and the resulting output
modes are subsequently surveyed by three identical photodetector. The
measured photocurrents are proportional to $\hat I_n$, $n=1,2,3$ given by
\begin{eqnarray}
\hat I_n &=& b^{\dag}_n b_n = \frac{1}{3} \sum_{k,l=1}^3
\exp\left\{i\theta_n (l-k)\right\} a_k^{\dag} a_l\;, \qquad
\theta_n=\frac{2\pi}{3}(n-1)
\label{pht}\; .
\end{eqnarray}
After photodetection a Fourier transform (FT) on the photocurrents is 
performed
\begin{equation}
\hat {\cal I}_s \equiv {\rm FT}(\hat I_1,\hat I_2,\hat I_3) 
= \frac{1}{\sqrt{3}} \sum_{n=1}^3 \hat I_n
\exp\left\{-i\theta_n (s-1)\right\}\; , \qquad s=1,2,3 \, .
\label{FT1}
\end{equation}
This procedure is
a straightforward generalization of the customary two-mode balanced
homodyning technique. 
By means of the identity
\begin{equation}
\delta_3 (s-1)
=
\frac{1}{3}\sum_{n=1}^3 \exp\left\{i\frac{2\pi}{3}n(s-1)\right\}
\label{FT2}\;,
\end{equation}
for the periodic (modulus 3) Kronecker delta  $\delta_3 $, we  
obtain our final expressions for the Fourier transformed photocurrents
\begin{eqnarray}
\hat {\cal I}_1 &=& 
\frac{1}{\sqrt{3}} \left\{
a^{\dag}_1 a_1 + a^{\dag}_2 a_2 + a^{\dag}_3 a_3
\right\} \; ,
\\
\hat {\cal I}_2 &=&  
\frac{1}{\sqrt{3}} \left\{
a^{\dag}_1 a_2 + a^{\dag}_2 a_3 + a^{\dag}_3 a_1
\right\} \; ,
\\
\mbox{and}\quad \hat {\cal I}_3 &=& 
\frac{1}{\sqrt{3}} \left\{
a^{\dag}_1 a_3 +a^{\dag}_2 a_1+ a^{\dag}_3 a_2
\right\}
\label{FT3}\;.
\end{eqnarray}
$\hat {\cal I}_1 $ gives no relevant information as it is insensitive to the phase
of the signal field, whereas $\hat {\cal I}_2 $ and $\hat {\cal I}_3$
are hermitian conjugates of each other and contain the relevant information
in their real and imaginary part. 
\par
In the following let us assume $a_1$ is the signal mode and $a_2$
is fed by a highly excited coherent state $| z \rangle$ representing 
the local oscillator. Since the local oscillator serves as the reference phase
($\varphi = 0)$, we have $| z \rangle =| |z| \rangle$.
For large $z$ the output photocurrents are intense enough to be 
easily detected and can be combined to give the reduced photocurrents
\begin{eqnarray}
\hat {\cal Y}_1 &=& \sqrt{3}\frac{\hat {\cal I}_2 + \hat {\cal I}_3 }{2 |z|} =
\hat a_{1} ( 0) + \hat a_{3}( 0 ) + O[\frac{1}{|z|}] \nonumber \\
\hat {\cal Y}_2 &=& \sqrt{3}\frac{\hat {\cal I}_2 - \hat {\cal I}_3 }{2 i|z|}=
\hat a_{1} (-\pi/2) - \hat a_{3} (-\pi/2) + O[\frac{1}{|z|}]
\label{THP}\;,
\end{eqnarray}
which we refer to as the {\em triple homodyne photocurrents}.
In Eq.~(\ref{THP}) $\hat a (\varphi) = 1/2 (a^{\dag}e^{i\varphi} +
ae^{-i\varphi} )$ denotes a quadrature operator of the field. 
\par
Each experimental outcome from a trit\-ter homo\-dyne detec\-tor is a pair
$(y_1,y_2)$ of real numbers from the joint measurements of
$\hat {\cal Y}_1 $ and $\hat {\cal Y}_2$. The 
corresponding two-dimen\-sio\-nal probability distri\-bution
of such outcomes is given by
\begin{equation}
P(y_1,y_2) = \int_{\bf R} \frac{d\lambda_1}{2\pi}
\int_{\bf R} \frac{d\lambda_2}{2\pi} e^{-i\lambda_1 y_1
-i\lambda_2 y_2}\; \Xi (\lambda_1,\lambda_2)
\label{FTP}\;,
\end{equation}
which is a double Fourier transform of the 
characteristic function $\Xi (\lambda_1,\lambda_2)$
of the device, namely
\begin{equation}
\Xi (\lambda_1,\lambda_2) = \hbox{Tr}\left\{ (\hat\rho_1 \otimes \hat\rho_3) \;
e^{i\lambda_1 \hat{\cal Y}_1+i\lambda_2 \hat{\cal Y}_2} \right\}
\label{Ch1}\;.
\end{equation}
Inserting Eq.~(\ref{THP}) in Eq.~(\ref{Ch1}), the characteristic
function can be rewritten in terms of the complex variable
$\gamma = (\lambda_2 + i\lambda_1)/2 $ yielding 
\begin{eqnarray}
\Xi (\gamma ,\bar{\gamma}) &=& \hbox{Tr}\left\{( \hat\rho_1 \otimes \hat\rho_3 ) \;
e^{\gamma a_1^{\dag}-\bar{\gamma}a_1} \; \; 
e^{-\gamma a_3^{\dag}+\bar{\gamma}a_3} \right\} \nonumber \\
&=&
\chi_{0|1}(\gamma,\bar{\gamma}) \chi_{0|3}(-\gamma,-\bar{\gamma})
\label{Ch2}\; .
\end{eqnarray}
It factorizes into two characteristic functions $\chi_0$,
see Eq.~(\ref{Xis}), one for each mode.
This and Eq.~(\ref{FTP}) -- changed to variables $\gamma$ -- gives $P$ as
an integral over the complex plane 
\begin{equation}
P(y_1,y_2) = \int_{\bf C} \frac{d^2\gamma}{\pi^2}
e^{\bar{\gamma} (y_1-iy_2)-\gamma (y_1+iy_2)}
\chi_{0|1}(\gamma,\bar{\gamma}) \chi_{0|3}(-\gamma,-\bar{\gamma})
\label{FTg}\; .
\end{equation}
We find that $P$ is a convolution integral reminiscent of $K_{SP}$ in 
Eq.~(\ref{Con}),
together with Eq.~(\ref{Wdf}) this leads to our {\em main result}:
\begin{equation}
P(y_1,y_2) = K_{13} (y_1-iy_2, y_1+iy_2 )
\label{Res}\; .
\end{equation}
It shows that the outcome probability distribution
from a triple coupler homodyne detection gives the desired
phase space density of Eqs.~(\ref{Con}) and (\ref{Tra}),
the mode $a_3$ represents the probe mode $P$ whereas $a_1$ is the signal mode $S$. 
When the probe is left in
the vacuum state, Eq.~(\ref{Ch2}) reads
\begin{eqnarray}
\Xi (\gamma ,\bar{\gamma}) &=&
\hbox{Tr}\left\{ \hat\rho_1 \otimes |0\rangle\langle 0|\;
e^{\gamma a_1^{\dag}-\bar{\gamma}a_1} e^{-\gamma a_3^{\dag}+\bar{\gamma}a_3}
\right\} = \nonumber \\
&=& \chi_{0|1}(\gamma,\bar{\gamma})
e^{-\frac{1}{2}|\gamma |^2} = \chi_{-1|1}(\gamma,\bar{\gamma})
\label{Ch3}\;,
\end{eqnarray}
namely, the triple coupler homodyning device becomes a Husimi $Q$-function detector. 
Note that customary eight-port homodyne thus mimics this scheme with one more input
and output.
\par
We end the paper with Fig.~\ref{f:EX1}, where we illustrate some examples
of phase space ityropensities accessible by triple coupler homodyne detection.
A coherent state of real amplitude $\beta =1$ is chosen for the signal:
Feeding vacuum into the probe mode $a_3$ the Husimi
$Q$-function $K_{|\beta\rangle |0 \rangle}(\alpha,\bar{\alpha}) =
Q(\alpha,\bar{\alpha}) = 1/\pi\exp\{-|\alpha-\beta|^2\}$
is obtained,  
if a squeezed vacuum $|0,r\rangle$ is used instead,
the distribution $K_{|\beta\rangle |0,r\rangle}(\alpha,\bar{\alpha}) =
1/\pi |\langle \alpha,r|\beta\rangle|^2$ results, and 
a number state $| n\rangle$ gives the distribution
$K_{|\beta\rangle | n\rangle}(\alpha,\bar{\alpha}) = 1/\pi |\langle n|\hat D^{\dag}(\alpha)|
\beta\rangle|^2$. 
\par
The authors are grateful to Prof. H. Paul for his hospitality in
the group ``Nichtklassische Strahlung" of the Max-Planck Society.
M.G.A.P. is partially supported by the University
of Milano with a postgraduate grant. A.V.C. and O.S. acknowledge
the support of the Max-Planck Society.

\newpage
\begin{figure}[h]
\caption{Schematic diagram of a triple coupler and its realization in terms of
50:50 beam splitters (BS) and phase shifters '$\varphi$'. In order to obtain a
symmetric coupler the following values has to be chosen: $\varphi_1=\arccos
(1/3)$} and $\varphi_2 = \varphi_1 /2$.
\label{f:TTT}
\end{figure}
\begin{figure}[h]
\caption{Outline of triple coupler homodyne detectors:
The hexagonal box symbolizes the 
electronically performed Fourier transform (FT).}
\label{f:THD}
\end{figure}
\begin{figure}[h]
\caption{Examples of phase space densities measurable by the triple coupler 
homodyne detection. For a coherent state of amplitude
$\beta =1$ we find (a) the Husimi $Q$-function obtained if the
probe mode $a_3$ is the vacuum, (b) the distribution which results
if the probe is a squeezed vacuum $|0,r\rangle$ with the mean photon
number $\sinh^2 r= 1$, and (c) the distribution which results if
the probe is a number state $| n\rangle$ with $n= 1$.}
\label{f:EX1}
\end{figure}

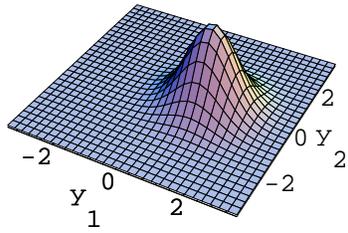
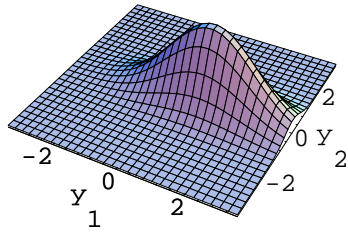
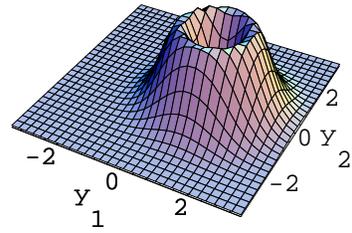
\end{document}